\documentclass{article}%
\usepackage{amssymb}
\usepackage{amsmath}
\usepackage{amsfonts}
\usepackage{graphicx}%
\begin{document}

\begin{center}
{\Large Critical temperatures in cuprate superconductors }
\end{center}

\[
\]
M. Salis

\textit{Department of Physics, University of Cagliari, s.p. 8 Km 0.7, 09042 -
Monserrato, Cagliari, Italy}

CGS\textit{\ Laboratory, University of Cagliari, s.p. 8 Km 0.7, 09042 -
Monserrato, Cagliari, Italy}

*Electronic address: masalis@unica.it%
\[
\]

\[
\]

\textbf{Abstract.- }The order parameter in superconductivity of cuprates is
investigated in the framework of the Bogoliubov theory. By using a simplifying
assumption about the electronic states, it is predicted an effective critical
temperature $T_{c}^{\ast}>T_{c}$ associated to the coherent gap $\Delta_{0}$.
A connection between $\Delta_{0}$ and the antinodal pseudogap $\Delta_{PG}$ is
proposed allowing for a comprehensive picture of the main experimental
features of cuprates superconductors.%

\[
\]

The high temperature superconductivity (SC) in cuprate based materials is one
of the most intensely investigated topics in condensed matter physics. It has
been about twenty-four years since the seminal paper of Bednorz and Muller [1]
sparked the race to attain the highest \ SC critical temperature($T_{c}$) [2].
Much efforts have been spent to explain the pairing nature of electrons in the
SC phase of cuprates, but yet physicists remain undecided on what model among
the ones so far proposed worths general consensus [3]. However, in recent
years several features concerning the SC and the normal state have been
unveiled so making the matter a little more clear. It is known that, due to
high electronic repulsion, undoped $CuO_{2}$ planes are Mott insulators turned
to antiferromagnetic (AF) states [3]. Injection of small quantities of holes
destroys\ the long-range AF order and give rise to phenomenologically rich
electronic systems. Below a temperature T*, depending on the doping level, the
normal phase show a gap of width dependent on the electron momentum
(pseudogap, PG). As the doping level increases a small doping interval
initially is found where the electronic system shows a glassy phase with
partial electron and hole localization [4,5]. Further increasing the doping,
open Fermi contours appear across the nodal directions in the momentum space,
usually referred to as Fermi arcs [6].

By means of ARPES measurements, it was found that the SC gap exhibits the
universal form $\Delta_{0}~\cos2\theta_{k}$ over an angular range larger than
the Fermi arc [7]. Beyond this, the spectral weight of the SC states decreases
against the one of the PG states [7]. Experiments also show that $T_{c}$ is
approximately related both to $\Delta_{0}$ and to the extension of Fermi arcs
by [8,9]
\begin{equation}
2\Delta_{0}~\alpha/(\pi/4)=4.3~k_{B}T_{c}~\ , \label{nqX0}%
\end{equation}
where the angle $\alpha$ spans the\ half-Fermi arc as measured from the nodal
direction [8]. Thus, the bell-shaped curve of $T_{c}$ versus the hole
concentration is strongly dependent on the way the shrinking PG hamper the
full expression of the SC gap and the Fermi contour. On this regard, it is
worth to point out that available data does not allow to find an unambiguous
connection between the competing gaps, because of their very different
behaviors [11,12].

In this paper, some points concerning the order parameter are reconsidered on
the light of the Bogoliubov theory. By using a simplifying assumption about
the electronic state distribution, worthy for semi-quantitative discussion, an
effective critical temperature $T_{c}^{\ast}>T_{c}$ is predicted associated to
the coherent gap $\Delta_{0}$. A connection is proposed between $\Delta_{0}$
and the antinodal pseudogap $\Delta_{PG}$, consistent with experimental
findings in cuprates superconductors. At the end of this paper, a formal
construction of the anisotropic pairing matrix is proposed.

According to the Bogoliubov theory, the order parameter satisfies [13,14]%

\begin{equation}
\Delta_{k}=\frac{1}{2}\sum_{k^{\prime}}\frac{G_{k,k^{\prime}}\Delta
_{k^{\prime}}}{\sqrt{\xi_{k^{\prime}}^{2}+\Delta_{k^{\prime}}^{2}}}~\ ,
\label{nqX1}%
\end{equation}
where $\xi_{k}$ are the energies of quasi-particle states with respect to the
Fermi level. Here it is used the convention according to which the minus sign
of the pairing interaction is explicitated in the Bogoliubov Hamiltonian [14].
In agreement with the d-like form of $\Delta_{k}$, we will use $G_{k,k^{\prime
}}=G~\cos2\theta_{k}~\cos2\theta_{k^{\prime}}$ where, for simplicity, $G$ is
assumed to be independent of $k$ and $k^{\prime}$. This position will be
discussed later in more detail. Simple calculus leads to%

\begin{equation}
2\Delta_{0}=3.52~\Gamma_{0}~kT_{c}~\ , \label{nqX2}%
\end{equation}
where for the case of closed Fermi surface $\Gamma=1.22$ [15].

Differently from the usual approach [15], a reduced angular range will be
considered in the sum (\ref{nqX1}). The reason is that, because of competition
with PG states, when approaching the antinodal direction, there may be not
enough states allowing for coherent pairing [16]. Thus an effective angle
$\alpha^{\ast}$ is here introduced, not necessarily coincident with the
angular extension $\alpha$ of the half-Fermi arc. On this ground, eq.
(\ref{nqX2}) is replaced by%

\begin{equation}
2\Delta_{0}=3.52~\Gamma\left(  \alpha^{\ast}\right)  ~kT_{c}^{\ast}~\ ,
\label{nqX4}%
\end{equation}
where $\Gamma\left(  \alpha^{\ast}\right)  \geq\Gamma_{0}$ (\ $\Gamma\left(
\pi/4\right)  =\Gamma_{0}$). Thus, $T_{c}^{\ast}$ is to be distinguished from
$T_{c}$ since the former pertains to the closure of gap $\Delta_{0}$ [11] and
the latter to the drop of the superfluid density [7]. Actually, there are
spectroscopic evidences of presence of coherent pairing above $T_{c}$
$[8,16,17]$. Loss of phase rigidity $[17]$ or gap nucleation in nanoscale
regions [11] are invoked to explain the signals of the phase-incoherent
superconductivity.\ However, besides the proposed interpretations, it is
reasonable to assume that the vanishing of the residual SC phase is signed by
$T_{c}^{\ast}$, like the main drop of the phase-coherent superconductivity is
signed by $T_{c}$.

To get into more details, it is convenient to use a simplifying model for the
electron states.\ For this purpose, let us assume that states contributing to
pairing lie within an energy span $\pm\delta\varepsilon$ around the Fermi such
that $\delta\varepsilon<<$ $kT_{c}^{\ast}$. If $\Omega_{0}$ is the density of
states at the Fermi level, eq. (\ref{nqX1}) leads to%

\begin{equation}
\Delta_{0}=8\Omega_{0}G\delta\varepsilon\sin^{2}\alpha^{\ast}~\ \ ,
\label{nq5}%
\end{equation}

\begin{equation}
3.52~\Theta\left(  \alpha^{\ast}\right)  ~kT_{c}^{\ast}=2\Delta_{0}~\ \ ,
\label{nq6}%
\end{equation}

\begin{equation}
\Theta\left(  \alpha^{\ast}\right)  =\frac{1}{3.52}\frac{4\sin^{2}\alpha
^{\ast}}{\frac{\alpha^{\ast}}{2}-\frac{1}{8}\sin4\alpha^{\ast}}~\ \ ,
\label{nq7}%
\end{equation}
A numerical check shows that $\Theta\left(  \alpha^{\ast}\right)
/\Gamma\left(  \alpha^{\ast}\right)  \simeq1.2$ within the range $[0,\pi/4]$.
It is interesting that for small angles eqs (\ref{nq6}) and (\ref{nq7}) lead
to $(12/\pi)~kT_{c}^{\ast}\simeq2\Delta_{0}~\alpha^{\ast}/(\pi/4)$, which
mimics the equation expected for $T_{c}$ with $\alpha=\alpha^{\ast}$.\ 

In the model dealt-with, $\Delta_{0}$ is related to the pairing strength in a
very simple way. If PG and SC shares the same dependence on G, eq. (\ref{nq5})
insights what connection may be established between them. It was suggested, on
this regard, that competition may be originated by the same microscopic
interactions underlying the pairing mechanism [7]. Thus, the pairing strength
may enter in the particle self-consistent field to determine the pseudogap
amplitude $\Delta_{PG}$. Therefore, it is not meaningless to assume
$\Delta_{PG}\propto G$, so that
\begin{equation}
\Delta_{0}=\gamma\Delta_{PG}\sin^{2}\alpha^{\ast}~\ , \label{nq8}%
\end{equation}
where $\gamma$ is a suitable factor. Eventually, the latter could also include
adjustments required to account for interactions among CuO$_{2}$ planes [8].
Since in overdoped cuprates\ $\Delta_{0}$ can be larger than $\Delta_{PG}$ ,
it can be expected that $\gamma\gtrsim2$.

It may be helpful to consider some numerical examples by using data from ARPES
detailed investigations of Bi2201 superconductors [7]. The overdoped sample,
there labelled as OD29K, was found with $\Delta_{PG}\simeq\Delta_{0}%
\simeq14~meV$ and $T_{c}=29K$. The half coherence arc was estimated at about
37$%
{{}^\circ}%
$ measured from the nodal direction. By using this figure as $\alpha^{\ast}$,
$\gamma=2.75$ is obtained so that eq. (\ref{nq6}) leads to $T_{c}^{\ast}=59K$
($2\Delta_{0}/kT_{c}^{\ast}=5.6$). By using the same $\gamma$ for the
underdoped sample (UD23K, $T_{c}$=23K), showing $\Delta_{PG}\simeq62meV$ and
$\Delta_{0}\simeq19~meV$, we get $\alpha^{\ast}=19.2%
{{}^\circ}%
$ which is close to the estimated 20.4$%
{{}^\circ}%
$. In this case, $T_{c}^{\ast}=48.5~K$ ($2\Delta_{0}/kT_{c}^{\ast}=9.4$).
Discrepancy is obtained for the optimally doped sample (OP35K, $T_{c}=35K$),
showing $\Delta_{PG}\simeq34meV$ and $\Delta_{0}\simeq18~meV$, which allows
$\alpha^{\ast}=26%
{{}^\circ}%
$ against the estimated 30$%
{{}^\circ}%
$.

In place of any further comment, it is convenient to present a comprehensive
(but quite qualitative) picture of the obtained results by using the above
data. For this purpose, two hypothetical curves for the coherence and the
Fermi angles versus hole concentration, shown in fig. 1, are used in
eqs.(\ref{nqX0}),(\ref{nq5}-\ref{nq8}).\ Reference angles $\alpha$ are
obtained by means of eq. (\ref{nqX0}) by using data from ref [7]. In both
curves, the zero points are added as an assumption. In fig. 2, the curve for
$\Delta_{0}$ is calculated by using $\gamma=2.75$ and $\Delta_{PG}%
(meV)=75~(1-x)$ (which is a rough estimation for the material dealt with [8])
where x=0.05+(p-0.05)/0.225 (about the range where $T_{c}\neq0$ [8]), symbol p
meaning the hole concentration within [0.005,0.275]. For comparison, the
typical curve $T_{c}\varpropto x(1-x)$ is also drawn by circles. Finally, it
is to be noted that fig.2 appears quite similar to the analogous one reported
in ref. [8].

As for the interaction matrix, some insights are suggested by experiments
according to which pairing take place along the O-Cu-O lattice directions in
the $CuO_{2}$ plane [18]. Conjugated electrons may interact either along the X
or Y directions, reminiscent of the residual AF interaction. Thus,
probabilities can be associated to pairs in such a way that the mean
interaction matrix can be written as a tensor contraction like%

\begin{equation}
G_{k,k^{\prime}}=\sum_{\lambda_{1},\lambda_{2},,\lambda_{1}^{\prime}%
,\lambda_{2}^{\prime}}G_{k,k^{\prime}}^{\lambda_{1},\lambda_{2},,\lambda
_{1}^{\prime},\lambda_{2}^{\prime}}\rho^{\lambda_{1},\lambda_{2}}\rho
^{\prime\lambda_{1}^{\prime},\lambda_{2}^{\prime}}~, \label{nq2}%
\end{equation}
where $\rho$ and $\rho$' are suitable probability matrices of pairs. The more
simplest form for $\rho$, satisfying Tr($\rho$)=1, is the diagonal one%

\begin{equation}
\rho=%
\begin{pmatrix}
\cos^{2}\theta & 0\\
0 & \sin^{2}\theta
\end{pmatrix}
~\ \ ,\label{nq3}%
\end{equation}
For language convenience, let us define the mean polarization of the pair
along the direction $\theta$ as $P_{\theta}=\rho^{11}-\rho^{22}=\cos2\theta$.
Thus, polarizations $P_{\theta}=\pm1$ merely means pairs capable of
interaction along only one of the two main directions O-Cu-O. Nodal
directions, $P_{\theta}=0$, are the ones where polarization sign changes.
Based on these positions, we are allowed to use a 2x2 matrix $G_{k,k^{\prime}%
}^{\lambda,,\lambda^{\prime}}$ where meaning of elements can be easily
understood by considering pairs moving along the main directions. The diagonal
terms represent processes where incoming and outcoming pairs maintain the same
X or Y directions. Since these are expected to be equivalent, the relative
terms can be assumed to be equal. The non-diagonal terms correspond to
processes where the outcoming pairs are polarized on the direction orthogonal
with respect to the ones of the incoming pairs. Depending on the interaction
features, a phase factor may be associated to these terms so that, in a little
more general way, the anisotropic properties of the pairing matrix can be
represented as%

\[%
\begin{bmatrix}
G_{k,k^{\prime}}^{11} & G_{k,k^{\prime}}^{12}\\
G_{k,k^{\prime}}^{21} & G_{k,k^{\prime}}^{22}%
\end{bmatrix}
=g_{k,k^{\prime}}%
\begin{bmatrix}
1 & \exp(-i\varphi)\\
\exp(i\varphi) & 1
\end{bmatrix}
~\ ,
\]
where the s- and d-like cases correspond to $\varphi=0,\pm2\pi$ and
$\varphi=\pm\pi$, respectively.

In conclusion, by properly handling the Bogoliubov theory with some
simplifying assumptions, several points concerning the order parameter and its
relation with the critical temperatures can be accounted for. Hopefully, the
picture here presented may become an useful canvas for future investigations
in the concealed matter of high temperature superconductivity.%
\[
\]

\textbf{References}

[1] J. G. Bednorz and K. A. Muller, Z. Phys B \textbf{64}, 189(1986)

[2] G. F. Sun, K. W Wong, B. R. Xu, Y. Xin and D. F. Lu, Phys. Lett A
\textbf{192},122 (1994)

[3] D. A. Bonn, Nature Phys. \textbf{2}, 159 (2006)

[4] Y. Kohsaka, C. Taylor, P. Wahl, A. Schmidt, Jhinhwan Lee, K. Fujita, J. W.
Alldredge, K. McElroy, J. Lee, H. Eisaki, S. Uchida, D.-H. Lee, J. C. Davis ,
Nature \textbf{454} , 1072 (2008)

[5] Y. Kosaca, C. Taylor, K. Fujita, A. Schmidt, C. Lupien, T. Hanaguri, M.
Azuma, M. Takano, H. Eisaki, H. Takagi, S. Uchida, J. C. Davis, Science,
\textbf{315}, 1380 (2007)

[6] J. Meng, G. Liu, W. Zhang, L. Zhao, H. Liu, X. Jia, D. Mu, S. Liu, X.
Dong, J. Zhang, W. Lu, G. Wang, Y. Zhou, Y. Zhu, X. Wang, Z. Xu, C. Chen, X.
J. Zhou Nature \textbf{462}, 335 (2009)

[7] . T. Kondo, R. Khasanov, T. Takeuchi, J. Schmalian and A Kaminski, Nature
\textbf{457}, 296 (2009)

[8] T. Yoshida, M. Hashimoto, S. Ideta, A. Fujimori, K. Tanaka, N. Mannella,
Z. Hussain, Z-X. Shen, M. Kubota, K. Ono, S. Komiya, Y. Ando, H. Eisaki and S.
Uchida, Phys. Rev. Lett. \textbf{103}, 037004 (2009)

[9] A. Pushp, C. V. Parker, A. N. Pasupathy, K. K. Gomes,S: Ono, J. Wen, Z.
Xu, G. Gu, A. Yazdani, Science \textbf{324},1689 (2009)

[10] K. Tanaka, W. S. Lee, D. H. Lu, A. Fujimori, T. Fujii, Risdiana, I.
Terasaki, D. J. Scalapino, T. P. Deveraux, Z. Hussain, Z-X Shen, Science ,
\textbf{314 }, 1910 (2006)

[11] K. K. Gomes,A. N. Pasupathy, A. Pushp, S. Ono, Y. Ando, A. Yazdani,
Nature \textbf{447}, 569 (2007)

[12] W. S. Lee , I. M. Vishik, K. Tanaka, D. H. Lu, T. Sasagawa, N. Nagaosa,
T. P. Deveraux, Z. Ussain, Z. X. Shen, Nature \textbf{450}, 81 (2007)

[13] G. Grosso and P. Parravicini , Solid State Physics (Academic Press, San
Diego, San Francisco, New York, Boston, London, Sydney, Tokyo, Toronto, 2000).
Ch. 18

[14] S. T. Belyaev, Effect of pairing correlations on nuclear properties, Mat.
Fys. Medd. Dan. Vid. Selsk, \textbf{31}, no 11 (1959)

[15] H. Won, K. Maki, Phys. Rev. B \textbf{49}, pp 1397(1994)

[16] T. Kondo, T. Takeuchi, A Kaminski, S. Tsuda, S. Shin, Phys. Rev. Lett.
\textbf{98}, 267004 (2007)

[17] J. Lee, K. Fujita, A. R. Schmidt, C. K. Kim, H. Eisaki, S. Uchida, J. C.
Davis Science \textbf{325}, 1099 (2009)

[18] H. B. Yang, J. D. Rameau, P. D. Johnson, T. Valla, A. Tsvelik and G. D.
Gu, Nature \textbf{456}, 77(2008)

[19] L. D. Landau, E. M. Lifshitz , Quantum Mechanics (Pergamon Press, Oxford,
London, Edinburgh, New York, Paris, Frankfurt ,1965) Ch. 8%

\[
\]

\textbf{Captions}%

\[
\]
Figure 1.- Models of coherence and Fermi angle curves versus the hole doping
concentration. Triangles and circles are obtained by using data from ref. [7]
in eq. 8 and in eq. (1), respectively. The zero points an assumption.%

\[
\]
Figure 2.- Pseudogap ($\Delta_{PG}$), coherent gap ($\Delta_{0}$) and critical
temperature curves versus the hole doping concentration: curve of $\Delta_{0}$
is obtained form eq. (8); curve of $T_{c}^{\ast}$ is obtained from eqs. (6)
and (7); curve of $T_{c}$ is obtained from eqs. (1). Curve drawn by circles
shows the typical form $T_{c}\varpropto x(1-x)$ where x stands for the
normalized hole concentration (see text).

\end{document}